\pdfoutput=1
\documentclass{article}
\usepackage{microtype}
\usepackage{graphicx}
\usepackage{subcaption}
\usepackage{booktabs}
\usepackage{hyperref}

\usepackage[preprint]{icml2026}
\hypersetup{pdfsubject={}}  %

\usepackage[utf8]{inputenc}
\usepackage[T1]{fontenc}
\usepackage{xurl}      %
\usepackage{placeins}  %
\usepackage{float}     %
\usepackage{amsmath,amsfonts,amssymb}
\usepackage{multirow}
\usepackage{array}
\usepackage{colortbl}
\usepackage{xspace}

\definecolor{memOr}{RGB}{213,94,0}   %
\definecolor{memBl}{RGB}{0,114,178}  %

\newcommand{\rdel}[1]{}

\newcommand{\mdel}[1]{}
\newcommand{\tjnote}[1]{}
\newcommand{\jmnote}[1]{}

\newcommand{\Gen}{MixAudio\xspace}
\newcommand{\Det}{musicDNA\xspace}

\newcommand{\demourl}{https://neutune.github.io/attr2027demo/}  %

\graphicspath{{figures/}}

\icmltitlerunning{Tracing Inputs, Verifying Outputs: Validating Attribution in Music Generation}

\begin{document}

\twocolumn[
\icmltitle{Tracing Inputs, Verifying Outputs: \\ Validating Attribution in Music Generation}

\begin{icmlauthorlist}
\icmlauthor{Taejun Kim}{neu}
\icmlauthor{Wonil Kim}{neu}
\icmlauthor{Jongmin Jung}{neu}
\icmlauthor{Hyeongseok Wi}{neu}
\icmlauthor{Sangeun Kum}{neu}
\icmlauthor{Keunhyoung Luke Kim}{neu}
\icmlauthor{Taehyoung Kim}{neu}
\icmlauthor{Dongjoo Moon}{neu}
\icmlauthor{Seungsoon Park}{neu}
\icmlauthor{Taewan Kim}{neu}
\icmlauthor{Virginie Berger}{neu}
\icmlauthor{Juhan Nam}{kaist}
\icmlauthor{Jongpil Lee}{neu}
\end{icmlauthorlist}
\icmlaffiliation{neu}{Neutune, Seoul, South Korea}
\icmlaffiliation{kaist}{KAIST, Daejeon, South Korea}
\icmlcorrespondingauthor{Taejun Kim}{taejun@neutune.com}  %
\icmlkeywords{music generation, attribution, memorization, provenance}

\vskip 0.3in
]

\printAffiliationsAndNotice{}  %

\begin{abstract}
How can we verify whose music contributed to an AI-generated output?
This paper demonstrates how input-based attribution can provide verifiable evidence of which audio sources were used in a generation and whether they shaped the output. To do so, we condition the generation solely on audio without any text input, then trace the inputs behind each output, and establish their musical effect. In prompt adherence tests and controlled input swaps, the stems generated by our generator, \emph{\Gen}, follow the prompt audio in timbre and the context audio in harmony. Yet these outputs may still reproduce training data not supplied as inputs. We therefore audit memorization with our musical version identification model, \emph{\Det}, and find few reproductions outside the input records. On human-judged cases within the flagged pool, it achieves higher precision and recall than the other tested memorization detectors. The two evaluations suggest that input records and output analysis provide complementary evidence for attribution, on which rights-holder reporting and compensation can draw as the AI music economy takes shape.
Audio examples are available at \url{\demourl}.
\end{abstract}

\section{Introduction}
\label{sec:intro}

In music generation, models use patterns learned from existing music to fulfill users' requests for particular structures or playing styles. This process has led to copyright disputes, including the recent \citet{gema2026suno} case. In a judgment not yet final, the court of first instance held that protected works memorized by the model and reproduced in its outputs infringed copyright. Beyond the issue of reproduction, the case raises a fundamental question for attribution: what actually counts as an existing work’s contribution to a generated output?

In this paper, we define attribution for a generated output as identifying which audio sources were used as inputs and verifying whether their musical characteristics shaped it. Once the evidence supports these claims and its scope is clear, it can ground recognition of the identified source's contribution and inform rights-holder reporting and compensation.

Attribution is difficult because it requires separate evidence for two claims: a particular track was used in generation, and its musical characteristics shaped the output. Resemblance in rhythm, harmony, or playing style cannot confirm the track's use as an input, since these characteristics may also occur in other tracks in the training data. In turn, confirmed input use does not establish that the track’s musical characteristics shaped the output.

We resolve the first uncertainty by conditioning generation on audio with known provenance and recording each input~\citep{kim2025generation, morreale2025abd}. This identifies which audio sources were used without inferring their use from the output. We then evaluate whether those inputs shaped the generated music, grounding attribution in both the record and evidence of musical effect.

Even when conditioning is faithful, a generated output may still reproduce a training track that was not supplied as input~\citep{roh2025confetti}. The input record cannot account for the relationship between that track and the output, so we compare generated outputs with the training data separately.

To implement and evaluate this approach, we present \emph{\Gen} and \emph{\Det}. \Gen, the generator,  produces music from audio with known provenance and records which inputs are used by stem and track section. \Det, a version identification model, compares generated stems with training data to find potential reproductions of works not supplied as inputs.

Conditioning tests show that \Gen's generated stems reflect the intended musical characteristics of their stem-specific inputs and the accompanying music. An \Det-based audit detected rare reproductions outside the input records. In a separate comparison on the stems flagged by any retriever, it identified confirmed reproductions more reliably than the other methods tested. These findings support our approach to music attribution, which combines input records and output analysis as complementary evidence.

\begin{figure*}[t]  %
  \centering
  \includegraphics[width=0.9\textwidth]{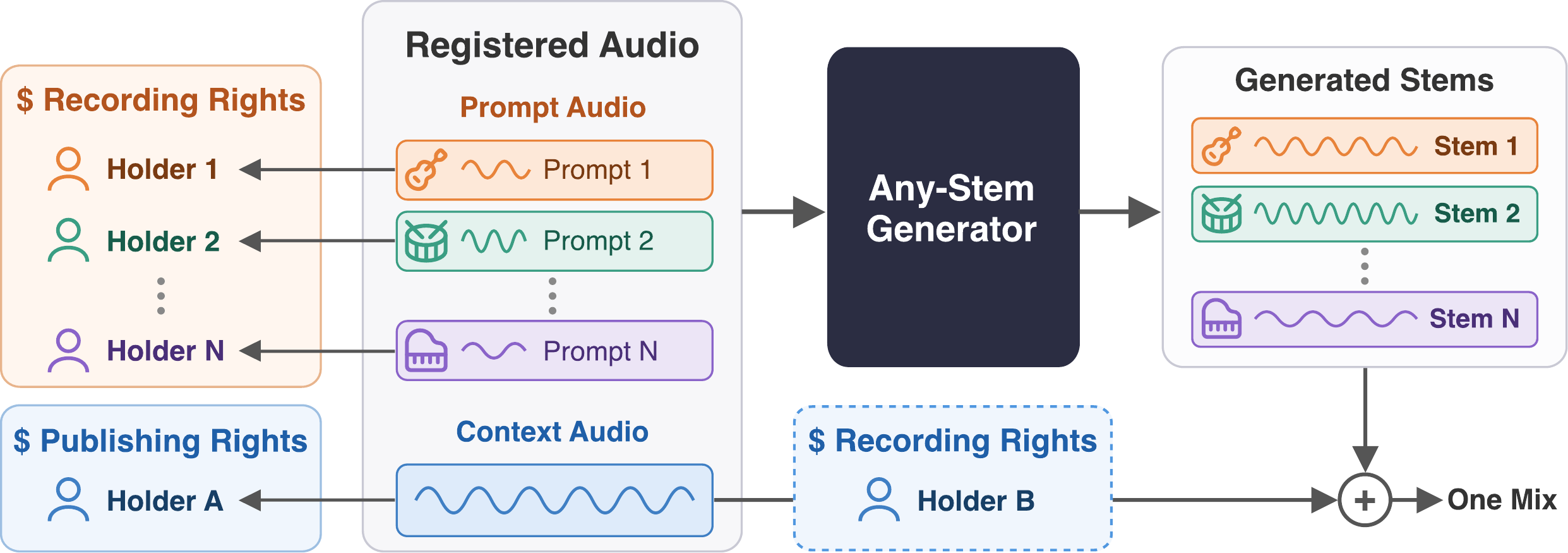}
  \caption{\textbf{Attribution by conditioning provenance.} A single
  pass generates \emph{any} subset of a track's stems, conditioned
  exclusively on registered audio, with no text input: one prompt
  recording per stem and an optional shared context. The two rights panels on
  the left are the rights involved at generation. The dashed box is
  added when the context is mixed into the released master.}
  \label{fig:attribution}
\end{figure*}

\section{Background and Related Work}
\label{sec:related}

\subsection{\texorpdfstring{ Approaches to Attribution}{Approaches to Attribution}}
\label{sec:related:attribution}

Prior research offers three approaches to attribution. Each uses different evidence to assess music use and whether its characteristics shaped the generated output. We distinguish the claims each approach can support.

Training data attribution (TDA) uses influence estimates to support claims about a training work's contribution to an output~\citep{koh2017influence,park2023trak,hammoudeh2024survey}. In music generation, it has been used to attribute outputs to training tracks~\citep{choi2025unlearning} and proposed as a basis for royalties~\citep{deng2025royalty}. The estimates are not yet reliable enough for that role~\citep{mlodozeniec2025influence}, because they approximate what removing a work from training would change, and that quantity can be verified only by retraining, which is rarely feasible at scale. \citet{aria2025} find that the same training tracks can rank at the top across different outputs, so a score can reflect the scoring method rather than the specific output. Compensation schemes built on such scores are therefore sensitive to estimation noise~\citep{wcw2025credit}. Nonetheless, disclosing which works were used in training could support transparency and accountability, allowing compensation for that use to be considered separately.

Output-based attribution searches for training works that resemble a generated output~\citep{somepalli2023forgery,wang2023evaluating}.
\citet{barnett2024roots} retrieve similar training clips using audio embeddings and show that their similarity rankings largely agree with human listening judgments.
\citet{batlle2024mira} use multiple audio similarity measures to detect fragments of training music reproduced in generated audio.
These methods provide evidence of resemblance or reproduction in the output.
Although they cannot by themselves establish how a particular training work influenced the output~\citep{aria2025,deng2025royalty}, the approach remains useful for identifying reproduced works in outputs from any generation system~\citep{somepalli2023forgery,drep2024}.

Input-based attribution records which music was used in a particular generation. \citet{morreale2025abd} propose attribution-by-design, making the provenance of conditioning music part of the generation process. Applying this principle, \citet{kim2025generation} designed a music AI agent architecture that records component-level input use during generation for attribution. These records provide auditable evidence that  rights-holder reporting and compensation can draw on. This paper evaluates the step that work left open: whether those recorded inputs actually shape the output as intended.

\subsection{Audio Conditioning and Faithfulness}
\label{sec:related:validation}

The evaluation requires defining which musical characteristics each input is intended to influence. Prior studies illustrate uses of audio conditioning. \citet{rouard2024audioconditioning}, \citet{nistal2024diffariff}, and \citet{yuan2026yue} use a short audio reference to condition the style of generated music. \citet{parker2024stemgen}, \citet{nistal2024diffariff}, and \citet{karchkhadze2024mtmusicldm} generate new stems conditioned on existing stems. \Gen draws on these studies. A short reference recording, used as an audio prompt for each stem, is intended to guide timbre and playing style, while a mix of existing stems provides harmonic and rhythmic context.

Yet similarity alone cannot show whether an input shaped the output, since prompts from other tracks may be similarly close. We therefore compare each generated stem with its own prompt and prompts from other tracks. We also change the prompt or context while holding the other conditions fixed to test how each affects the output (Section~\ref{sec:eval:faithfulness}).

\subsection{Memorization Audits}
\label{sec:related:memaudit}

Conditioning tests address the effect of recorded inputs. Reproductions of other training works require a separate comparison with the training data.

Detecting reproduction depends on what kind of match is sought between an output and a training work. Six music generation systems that report memorization audits use different methods, including audio-token matches, general audio-embedding similarity, and representations designed to identify a work across different performances~\citep{dhariwal2020jukebox,agostinelli2023musiclm,copet2023musicgen,evans2024stableaudio2,evans2024stableaudioopen,yuan2026yue} (Appendix~\ref{app:audits}). Because these methods detect different forms of reproduction, a finding of no detected reproduction must be interpreted in light of what each method can detect.

We compare generated stems with training data using several methods, including \Det, to find candidate reproductions and verify them by listening (Section~\ref{sec:eval:memorization}). We then compare the methods' detection performance on human-confirmed cases to assess the reliability of the memorization audits (Section~\ref{sec:eval:detectors}).

\subsection{Stem-Level Input Records}
\label{sec:related:stemgen}

Different stems in a track may be generated from different reference recordings. Examining only the full mix makes it difficult to tell which input was used for each stem or which characteristics appeared in the result. To connect each reference recording with its effect on the output, its use must be recorded and the result evaluated at the stem level. MSDM and Multi-Track MusicLDM model relationships among stems~\citep{mariani2024msdm,karchkhadze2024mtmusicldm}, while Stemphonic allows flexible control over the number and configuration of generated stems~\citep{stemphonic2026}.

Producing separate stem outputs does not itself record which music was supplied as input. \Gen uses reference audio for each stem and, when provided, musical context shared across stems. It records reference use by stem and shared context by track section, linking each input to the output it conditions. Section~\ref{sec:method} details \Gen's architecture for stem-level generation and input-use recording.

\section{The Generator}
\label{sec:method}

\paragraph{Conditioning inputs.}
Figure~\ref{fig:attribution} shows the interface of the
model. The generator produces individual stems rather than a mixture,
and is conditioned only on registered audio.
The model is designed for two uses, (1) new-song generation and (2)
remix generation. (1) The user supplies prompt audio only and the model
generates a song from scratch, as most music generators do from a text
prompt. (2) The user modifies an existing track by adding, replacing, or
removing stems, and the model generates the stems that are added or
replaced. Here the model also receives a context, a mix of the track's
existing stems, or of those the user chooses to keep; the generated
stems are parts laid on top of it, meant to be played back together with
it. In both uses at least one prompt is always given, whereas the
context is given only in (2).  For attribution and rights accounting, the system associates
prompt use with the recording-side contribution of the reference
performance, and context use with the composition-side contribution that
shapes the harmony and tempo of the generated stem. This design choice
rests on the observed behavior of the model, evaluated in
Section~\ref{sec:eval:faithfulness}, not on copyright law. The context recording itself does enter the
system, but replacing the context leaves the timbre of the generated
stem with the prompt (0.92 to 0.89 in Table~\ref{tab:t3a}b), so we find
little trace of a recording-side contribution in the generated stems. When the context is mixed into the final master together
with the generated stems, the record also names the context's recording rights holder.

\paragraph{Any-stem generation.}
The model works on discrete tokens.
A neural audio codec with residual vector quantization turns every prompt, context, and target stem into a sequence of tokens, one per quantizer level and time step.
The generator predicts the tokens of the target stems; the token-matching tests of Section~\ref{sec:eval:detectors} operate on these tokens.
The unit of generation is a section of a track: the context is the section's existing mix, the targets are the section's new stems, and a track is assembled section by section.
The generator operates on a single input sequence that carries the prompts and the target stems to be generated; requesting more stems simply appends more prompt--target pairs to the sequence.
The context audio, when given, is prepended to this sequence; the context and the target stems share the same positional encoding, which keeps the generated stems temporally aligned with the context.
The section's musical key and beat positions are provided as additional inputs.
A stem-level positional encoding ties the $i$-th prompt to the $i$-th target, so the model knows which prompt belongs to which generated stem.

\begin{table*}[t]  %
\centering\small
\caption{\textbf{Prompt adherence.}
Similarity between generated stems and prompts.
\emph{Matched}: each stem against the prompt it was generated from.
\emph{Random}: the same generations scored against another track's prompt. $\Delta$ = Matched $-$ Random.
Bold: larger $\Delta$ per source.}
\label{tab:t1}
\begin{tabular}{llcccccc}
\toprule
 & & \multicolumn{3}{c}{MERT (Style)} & \multicolumn{3}{c}{MFCC (Timbre)} \\
\cmidrule(lr){3-5}\cmidrule(lr){6-8}
Source & System & Matched & Random & $\Delta$ & Matched & Random & $\Delta$ \\
\midrule
\multirow{2}{*}{Training} & \Gen & $0.550$ & $0.014$ & $\mathbf{0.535}$ & $0.914$ & $0.554$ & $\mathbf{0.360}$\\
 & ACE & $0.549$ & $0.121$ & $0.428$ & $0.838$ & $0.602$ & $0.236$\\
\arrayrulecolor{black!25}\specialrule{0.4pt}{2pt}{2pt}\arrayrulecolor{black}
\multirow{2}{*}{Validation} & \Gen & $0.516$ & $0.000$ & $\mathbf{0.516}$ & $0.911$ & $0.550$ & $\mathbf{0.361}$\\
 & ACE & $0.534$ & $0.123$ & $0.411$ & $0.831$ & $0.602$ & $0.229$\\
\arrayrulecolor{black!25}\specialrule{0.4pt}{2pt}{2pt}\arrayrulecolor{black}
\multirow{2}{*}{MoisesDB} & \Gen & $0.582$ & $0.124$ & $\mathbf{0.457}$ & $0.920$ & $0.599$ & $\mathbf{0.321}$\\
 & ACE & $0.652$ & $0.264$ & $0.388$ & $0.912$ & $0.674$ & $0.237$\\
\arrayrulecolor{black!25}\specialrule{0.4pt}{2pt}{2pt}\arrayrulecolor{black}
\multirow{2}{*}{MUSDB18} & \Gen & $0.599$ & $0.183$ & $\mathbf{0.416}$ & $0.942$ & $0.667$ & $\mathbf{0.276}$\\
 & ACE & $0.624$ & $0.313$ & $0.311$ & $0.942$ & $0.756$ & $0.186$\\
\bottomrule
\end{tabular}

\end{table*}

\begin{table*}[t]  %
\centering\small
\caption{\textbf{Conditioning swaps.}
Each value is the similarity between a generated stem and the \textbf{prompt} or \textbf{context} marked in its row.
\emph{Base}: the base generation.
\emph{Swap}: the regeneration after the swap; the other conditioning input is held fixed.
Averaged over the four dataset sources.}
\label{tab:t3a}
\begin{tabular}{llcccc}
\toprule
 & & \multicolumn{2}{c}{\Gen} & \multicolumn{2}{c}{ACE} \\
\cmidrule(lr){3-4}\cmidrule(lr){5-6}
Hypothesis & Metric & Base & Swap & Base & Swap \\
\midrule
\multicolumn{6}{l}{\emph{(a) Prompt replaced}}\\
\quad Timbre moves to the substituted \textbf{prompt} & MFCC & $0.63$ & $0.89$ & $0.69$ & $0.85$\\
\quad Timbre leaves the original \textbf{prompt} & MFCC & $0.92$ & $0.64$ & $0.88$ & $0.70$\\
\quad Harmony stays with the original \textbf{context} & Chroma & $0.20$ & $0.22$ & $0.10$ & $0.06$\\
\midrule\multicolumn{6}{l}{\emph{(b) Context replaced}}\\
\quad Harmony moves to the substituted \textbf{context} & Chroma & $0.00$ & $0.21$ & $0.00$ & $0.06$\\
\quad Harmony leaves the original \textbf{context} & Chroma & $0.21$ & $0.02$ & $0.10$ & $0.06$\\
\quad Timbre stays with the original \textbf{prompt} & MFCC & $0.92$ & $0.89$ & $0.88$ & $0.87$\\
\bottomrule
\addlinespace[2pt]
\multicolumn{6}{l}{\scriptsize Scale anchors (unrelated pair $\cdot$ matched reference): MFCC $0.66 \cdot 0.92$; Chroma $0.02 \cdot 0.20$.}
\end{tabular}

\end{table*}

\section{Evaluation}
\label{sec:eval}
We ask two things of the attribution record.
Each generated stem should follow the inputs named in the record, and generated stems should rarely reproduce training  stems whose rights holders are not named in the record.
Section~\ref{sec:eval:faithfulness} tests the first with conditioning experiments.
Section~\ref{sec:eval:memorization} tests the second with a memorization audit, and Section~\ref{sec:eval:detectors} uses the listeners' judgments to test memorization detectors.
Every similarity score in this section is read against a measured baseline, the level a system reaches when it does not follow its input.

\subsection{Setup}
\label{sec:eval:setup}

\begin{figure*}[t]  %
\centering
\includegraphics[width=.9\textwidth]{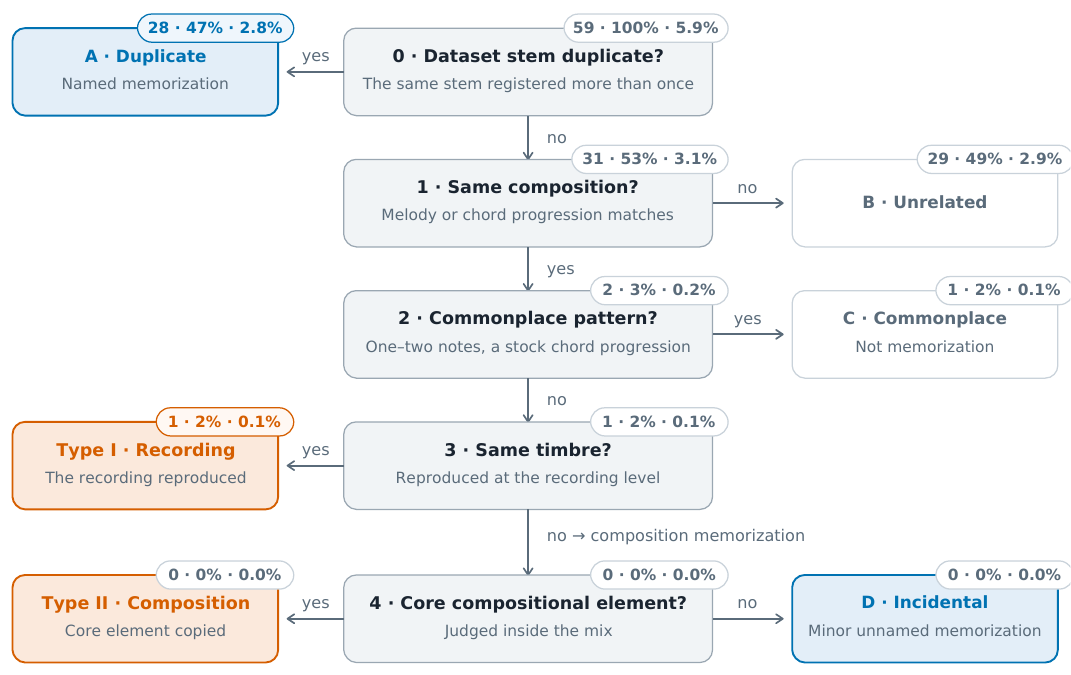}  %
\caption{\textbf{How listeners judge a flagged stem,} read top to bottom.
Orange: Type~I and II, the  material unnamed memorizations; blue:  named (A) or minor unnamed (D) memorization.
Badges: $n$ $\cdot$ \% of the band $\cdot$ \% of the  994 queries, shown for the \Det band over both splits (59 stems).
}
\label{fig:tree}
\end{figure*}

\paragraph{Conditioning sources.}
Prompts, contexts, beat positions, and keys come from four sources: the training and validation splits of the in-house multi-track dataset the model was trained on, and two external stem datasets, MoisesDB~\citep{pereira2023moisesdb} and MUSDB18~\citep{musdb18hq}.
The in-house dataset is licensed for both model training and service use.
The validation split is unseen but in-domain; the two external sets are unseen and out-of-domain.
 Vocal stems are never targets or prompts, so no lyrics condition a generated stem, though they can appear in a context mix.

\paragraph{Baseline.}
We compare with ACE-Step v1.5 in Lego mode~\citep{acestep2026}, ACE below.
It is the only system we know of that takes both an audio prompt and a context audio, conditions on the prompt audio itself, and generates at the stem level; other systems that accept both inputs first convert the prompt to text.
Table~\ref{tab:baselines} in Appendix~\ref{app:setup} lists the systems we reviewed.
We use its 2B checkpoint, the same size as \Gen, because the 4B checkpoint produced only artifact-laden output in Lego mode.
Lego adds one stem per call, so multi-stem instances are generated stem by stem, each new stem mixed into the context before the next is requested.

\paragraph{Instances.}
An instance is one section of a track.
Some of its stems are mixed into the context, and the system generates the rest.
For each stem to be generated, the prompt is an excerpt of that stem's own recording, so every generated stem has a real counterpart.
The generated stems are the ones the track actually has, so their number varies from instance to instance, and about a quarter of the instances have no context at all.
There are 6,810 instances and 15,034 target stems in total; the split rule, the excerpt rule, and the section length are given in Appendix~\ref{app:setup}.
Both systems generate the same instances from the same prompts and contexts, and generated stems that come out silent are dropped from every comparison.

\subsection{Conditioning Faithfulness}
\label{sec:eval:faithfulness}

\begin{table*}[t]  %
\centering\small
\caption{\textbf{Listener judgments on the flagged stems,} per retriever and split, in the categories of Figure~\ref{fig:tree}.
Percentages are shares of each column's flagged stems.
}
\label{tab:openset}
\setlength{\tabcolsep}{1.5pt}
\begin{tabular}{lrrrrrrrr}
\toprule
 & \multicolumn{4}{c}{Validation (unseen)} & \multicolumn{4}{c}{Training (registered)} \\
\cmidrule(lr){2-5}\cmidrule(lr){6-9}
Case & \Det & CLEWS & CLAP & MERT & \Det & CLEWS & CLAP & MERT \\
\midrule
Flagged & 25 & 25 & 25 & 25 & 34 & 35 & 31 & 24\\
\midrule
\textcolor{memOr}{Type I\,$\cdot$\,Recording} & 0 & 0 & 0 & 0 & \textbf{1}{\scriptsize(3\%)} & \textbf{1}{\scriptsize(3\%)} & 0 & 0\\
\textcolor{memOr}{Type II\,$\cdot$\,Composition} & 0 & 0 & 0 & 0 & 0 & 0 & 0 & 0\\
\midrule[.02em]
\textcolor{memBl}{A\,$\cdot$\,Duplicate} & 9{\scriptsize(36\%)} & 7{\scriptsize(28\%)} & 1{\scriptsize(4\%)} & 3{\scriptsize(12\%)} & 19{\scriptsize(56\%)} & 14{\scriptsize(40\%)} & 0 & 7{\scriptsize(29\%)}\\
B\,$\cdot$\,Unrelated & 16{\scriptsize(64\%)} & 12{\scriptsize(48\%)} & 24{\scriptsize(96\%)} & 22{\scriptsize(88\%)} & 13{\scriptsize(38\%)} & 16{\scriptsize(46\%)} & 31{\scriptsize(100\%)} & 17{\scriptsize(71\%)}\\
C\,$\cdot$\,Commonplace & 0 & 6{\scriptsize(24\%)} & 0 & 0 & 1{\scriptsize(3\%)} & 4{\scriptsize(11\%)} & 0 & 0\\
\textcolor{memBl}{D\,$\cdot$\,Incidental} & 0 & 0 & 0 & 0 & 0 & 0 & 0 & 0\\
\bottomrule
\end{tabular}

\end{table*}

The first requirement is that a generated stem follows its inputs: timbre and playing style from the prompt, and harmony from the context.
We test this in two ways.
Prompt adherence asks whether a generated stem resembles its own prompt more than another track's, and conditioning swaps ask whether replacing one input moves only the property it controls.

\paragraph{Prompt adherence.}
Table~\ref{tab:t1} scores each generated stem against its own prompt (Matched) and against another track's prompt (Random).
Style similarity is the cosine between MERT embeddings~\citep{li2023mert}, and timbre similarity is the cosine between MFCC vectors (Appendix~\ref{app:metrics}).
Adherence is usually reported as the Matched score alone~\citep{gong2025acestep,yuan2026yue}, and by that measure ACE looks better on style for most sources.
But ACE's Random scores are high as well: its outputs resemble one another, so an unrelated prompt also scores high.
The margin $\Delta$ = Matched $-$ Random removes this generic resemblance, and \Gen has the larger margin on every source and both metrics.

\paragraph{Conditioning swaps.}
Table~\ref{tab:t3a} asks what moves when one input is replaced.
We swap the prompt for another track's prompt (a), or the context for another track's context (b), and regenerate the stem. The context is swapped together with its key and beat inputs. The test asks whether the two inputs can be controlled independently, which needs both of them present, so only instances that have a context enter it.
Each row compares the new stem with one reference: the original or the substituted prompt, or the original or the substituted context.
Timbre is compared by MFCC similarity and harmony by bar-averaged chroma correlation (Appendix~\ref{app:metrics}).
Two anchors under the table set the scale: the similarity between unrelated clips, and the similarity between a stem and its own input.

Both systems follow the prompt's timbre.
In (a), similarity to the new prompt rises to the matched level (\Gen 0.63 to 0.89, ACE 0.69 to 0.85) and similarity to the old prompt falls (0.92 to 0.64, 0.88 to 0.70).
Harmony is where the two systems part.
In (b), \Gen's chroma correlation to the new context rises from 0.00 to 0.21, the matched level, and its correlation to the old context falls from 0.21 to 0.02, the unrelated level. Its timbre stays with the prompt (0.92 and 0.89).
ACE's correlation to the context stays between 0.00 and 0.10 in every condition, base generations included, so a swap has nothing to move.
Pitch-shifting or time-scaling an input instead of swapping it gives the same picture (Appendix~\ref{app:transform}).

\paragraph{Summary.}
Both mappings of Section~\ref{sec:method} hold. A generated stem resembles its own prompt rather than prompts in general, and the swaps separate the two inputs: replacing the prompt moves timbre and leaves harmony in place, and replacing the context moves harmony and leaves timbre in place. Each property therefore traces to one named input, which lets the record distinguish recording-side and composition-side contributions for rights accounting. ACE follows the prompt but not the context, so a record built on its outputs would support only the recording side. Generation quality is also higher than ACE's, in FAD on all three sources and in listener ratings (Appendix~\ref{app:quality}).

\subsection{Memorization Audit}
\label{sec:eval:memorization}

\begin{figure*}[t]  %
\centering
\includegraphics[width=.98\textwidth]{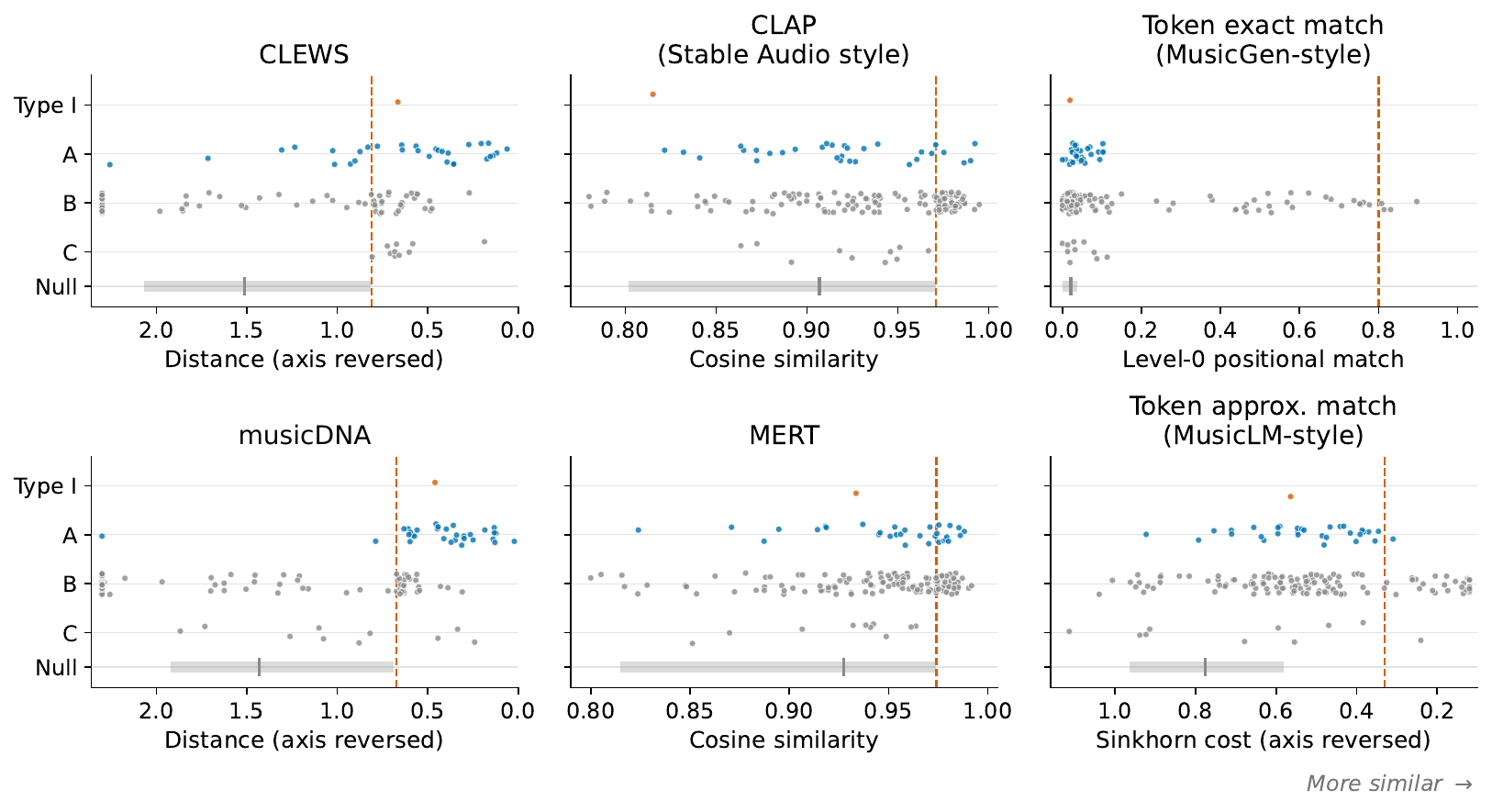}  %
\caption{\textbf{Detection performance of memorization metrics.}
One dot per human-judged stem, scored by all six detectors.
Rows follow the categories of Figure~\ref{fig:tree}, in the same colors: colored dots are confirmed memorizations (Type~I, A), grey dots are not (B, C).
More similar is to the right, and the dashed line is each detector's detection threshold.
The Null row (grey band) marks chance-level similarity: the 5--95\% range of scores for unseen validation stems.
Vertical jitter within a row carries no meaning.
}
\label{fig:five}
\end{figure*}

 Even when a generated stem follows its inputs, it can still reproduce a training stem it was never given as input.
The record names only the prompt, the context, and their rights holders, so a reproduced training stem has no entry in it.
The reproduction is  unnamed when the record does not name that stem's rights holders either.
The second requirement of Section~\ref{sec:eval} is that such unnamed reproductions are rare.  We test it by searching the training set with each generated stem and having listeners judge the closest matches.
Whether a match counts against the record depends on two terms, memorization and  named, which we define first.

\paragraph{Terms.}
A generated stem is a \emph{memorization} when it reproduces, in recording or in composition, a training stem other than its own conditioning inputs.
A memorization is \emph{ named} when the record already names the rights holders of the reproduced  stem, and \emph{ unnamed} when it does not.
Only an  unnamed memorization counts against the record.

Figure~\ref{fig:tree} shows the six categories a listener can assign.
 A \emph{duplicate} (A) is a case where the generated stem, the retrieved training stem, and the full stem the prompt was cut from all sound the same, because the retrieved stem is a second registration of the prompt's own recording.
The match could then come from the prompt as well as from memory, and listening cannot tell the two apart, so we count duplicates as memorizations, and as  named ones.
In every duplicate case the two registrations trace to the same rights holder. Most pairs arose when stems first delivered through an intermediary were delivered again under a direct contract with their producer, and the rest are repeat registrations within one catalog, so the rights holder named in the record is the same either way.
\emph{Type~I} is a reproduced recording: the generated stem matches the training stem in composition and in timbre.
\emph{Type~II} is a composition reproduced in a different timbre, where the copied element carries the piece.
Both are  unnamed, and we treat both as material.
\emph{D} is also a composition reproduced in a different timbre, but the copied element is incidental to the piece.
It is unnamed, and we treat it as minor.
\emph{B} and \emph{C} are not memorizations.
In B the two stems share no melody or chord progression, and in C they share only commonplace  elements, such as a stock chord progression.

\paragraph{Retrieval.}
The queries are the generated stems of melody-carrying instruments (guitars, keys, brass, and their relatives); bass and drum stems are not audited.
There are two splits of 497 queries each.
Validation queries are generated from the conditioning of validation tracks, which the model never saw.
Training queries reuse the exact conditioning a training track had during training (prompt, context, beat, and key), the setting most likely to draw memorization out.
Four retrievers search the training set with each query and score it by its nearest training stem: \Det, our version identification model, and CLEWS, a musical version matching model~\citep{serra2025clews}, both of which compare compositions, and CLAP~\citep{wu2023large} and MERT, which are general-purpose audio embeddings.
For a training query, the stems of its own conditioning track are left out of the search, since a generation that retrieves its own conditioning source is following its conditioning, and the record already covers that.
Each retriever's threshold is the 25th-highest score among its 497 validation queries, so it flags the top 5\% of the validation split, and the same threshold is applied to the training split (Appendix~\ref{app:retrieval}).

\paragraph{Listening.}
Three listeners, authors of this paper, place every flagged stem in one of the six categories (Appendix~\ref{app:listening}).
They hear three stems, each alone and again inside its context: the generated stem, the retrieved training stem, and the full-length stem the prompt was cut from.
Each stem is heard as a whole section, the unit of generation (Section~\ref{sec:method}).
Listeners cannot see which retriever flagged a stem. The three judgments are merged by majority vote (Appendix~\ref{app:listening}).
Whether a copied element is core (Type~II) or incidental (D) is judged inside the mix, against the piece as a whole.

\paragraph{Results.}
Table~\ref{tab:openset} counts the judgments per retriever and split.
 Unnamed memorization is rare:  one case, in the training split, a reproduction of a recording (Type~I) flagged by both \Det and CLEWS, and none in the validation split.
No Type~II or D case appears.
The other  31 of the 32 confirmed memorizations are duplicates.
Duplicate registration is a lapse in dataset curation rather than a property of the model, but it serves the audit: each duplicate case is a generated stem that closely matches a training recording, so duplicates supply most of the confirmed matches on which Section~\ref{sec:eval:detectors} tests the detectors. CLAP and MERT miss most of them.

\paragraph{Recurrence of the  Type~I case.}
To see how readily the  Type~I case produces  unnamed
memorization, we regenerated  it with different seeds.
 The case is regenerated eight times with its original conditioning, and a regeneration counts as a recurrence when its \Det distance to the retrieved stem stays under the flagging threshold.
 It recurs in all 8 seeds, presumably because that very conditioning was used in training. The context is what draws the memorized stem out. With the context removed, it does not recur (0 of 8). Just as lyrics, a
highly specific condition, drew training songs out of deployed
models~\citep{roh2025confetti}, the context audio is specific enough to
do the same.

 Within the scope of this audit, we found few reproductions of training stems outside the attribution record.

\subsection{Evaluating Memorization Detectors}
\label{sec:eval:detectors}

Because listeners judged every flagged stem, the audit leaves behind what previous audits lacked: a set of confirmed memorizations on which a detector can be scored (Section~\ref{sec:related:memaudit}).
We score six detectors on it, the four retrievers and the two token tests of previous audits.
Only the two composition-level retrievers find the confirmed memorizations.
The general-purpose audio embeddings and the token tests miss nearly all of them.

\paragraph{Setup.}
The judged pool is the 171 generated stems that any retriever flagged; it holds  32 confirmed memorizations (1 Type~I and 31 duplicates) and 139 stems that are not memorizations.
Recall is measured against the confirmed memorizations in this pool, so a memorization that no retriever flagged is not counted. \Det and CLEWS are two of the four retrievers that formed the pool, and the two token tests did not contribute to it.
Four detectors are the retrievers, with the flags of the audit.
The other two are the token tests of previous audits, run on our model's tokens.
The MusicGen-style test flags a pair when at least 80\% of their level-0 token positions match exactly, level-0 being the first and coarsest quantizer level.
The MusicLM-style test flags a pair when the transport cost between their token histograms is lower than the cost of all but 0.01\% of unseen validation stems.

\paragraph{Results.}
Figure~\ref{fig:five} places every judged stem on each detector's scale.
CLAP and MERT are driven by timbre: 96 to 100\% of CLAP's flags and 71 to 88\% of MERT's are unrelated stems (B), and CLAP finds one of the  32 confirmed memorizations.
The token tests fail in both directions.
The MusicGen criterion flags four stems, all unrelated, and the MusicLM-style test recovers  1 confirmed memorization while flagging 33 unrelated stems.
The composition-level retrievers do better, \Det best.
Of the  32 confirmed memorizations, \Det flags  29 and CLEWS  22, a recall of  0.91 against 0.69.
Of the stems each flags,  29 of \Det's 59 are confirmed and  22 of CLEWS's 60, a precision of  0.49 against 0.37.

\section{Discussion and Limitations}

\paragraph{No text conditioning.}
The model is conditioned on audio alone, so it cannot serve as a
text-to-music system, the dominant interface today. However, text can still enter
indirectly, by retrieving a registered prompt or context recording
through vector search. Audio is also the more precise control. Text maps
one-to-many to music, so text-conditioned outputs tend toward the average
of what fits a description, whereas a prompt recording fixes timbre and
playing style as concretely as the recording chosen. Whether audio-only
conditioning costs quality relative to text conditioning is not tested
here. Against the closest available baseline, which also takes an audio
prompt, generation quality is higher (Appendix~\ref{app:quality}).

\paragraph{Scope of the memorization audit.}
The audit finds that generations rarely reproduce training  stems the
attribution record does not name. This holds for this model on this
catalog, not for the architecture in general. Memorization depends on the model
and its training data, so a different catalog needs its own audit. The
audit also hears only flagged stems. A memorization that every retriever
scores below its threshold goes unexamined, so the counts of
Table~\ref{tab:openset} are lower bounds. The audit covers only
melody-carrying instruments, not bass or drum stems. The present catalog never uses vocal stems as targets or prompts, so no lyrics condition a generated stem
(Section~\ref{sec:eval:setup}). Lyrics are an unusually precise condition, and lyric prompts alone have drawn training songs out of deployed
models~\citep{roh2025confetti}. We expect vocal targets and lyric prompts to
produce composition-level (Type~II) cases that did not arise here.
Because no Type~II case arose, the detector comparison in
Section~\ref{sec:eval:detectors} does not test composition-level copying
in a different timbre. The reference labels of that comparison were set
by three listeners who are authors of this paper.

\section{Conclusion}

Attribution in music generation should identify which music was used and how its musical characteristics shaped the output. The reliability of this evidence is central to determining how rights holders are compensated. We examine what different approaches to attribution can establish, then empirically validate input-based attribution.

We evaluate two requirements for the attribution record by examining whether recorded inputs shape outputs as intended and whether outputs reproduce training tracks absent from the records.
\Gen's outputs reflected the intended musical characteristics of their inputs, and an \Det-based audit found that reproductions outside the input records were rare. On human-judged cases within the flagged pool, \Det achieved higher precision and recall than the other methods tested. These complementary findings provide an evidentiary basis which rights-holder reporting and compensation can draw on.

\section*{Impact Statement}\label{sec:impact}  %
This paper presents work whose goal is to advance the field of machine
learning. The attribution framework we propose can help direct revenue
from generated music to its rightful rights holders. As our results show,
however, it is not perfect. Its records should not be taken on trust, and
decisions of consequence should always involve human review. Beyond this,
there are many potential societal consequences of our work, none of which
we feel must be specifically highlighted here.

\section*{Acknowledgements}
This work was supported by the Korea Creative Content Agency (KOCCA) and
the Artificial Intelligence Industry Cluster Agency (AICA).

\label{sec:bib}\bibliography{refs}
\bibliographystyle{icml2026}

\clearpage
\appendix
\onecolumn  %

\section{Evaluation Setup Details}
\label{app:setup}

\begin{table}[H]
\centering\small
\caption{\textbf{Evaluation set composition.} One instance per
section; target stems are the stems each system generates; contextless
instances have no context mix. Hours: total section audio.}
\label{tab:composition}
\begin{tabular}{lcccccc}
\toprule
Source & Tracks & Sections & Instances & Target stems & Contextless & Hours \\
\midrule
Train & 1922 & 2073 & 2073 & 5030 & 515 & 7.1 \\
Val & 168 & 2086 & 2086 & 4960 & 518 & 6.9 \\
MoisesDB & 112 & 1058 & 1058 & 2266 & 282 & 5.6 \\
MUSDB18 & 150 & 1593 & 1593 & 2778 & 458 & 8.3 \\
\bottomrule
\end{tabular}

\end{table}

\paragraph{Sections.}
An instance is one section of a track. In-house tracks carry section
boundaries in their metadata; the external tracks are segmented with
All-In-One~\citep{kim2023allinone}, and each section is snapped to its first
and last downbeat. Sections shorter than 2\,s or longer than 40\,s are
excluded, as in training. The validation split is used in full. The
training split is a sample of about the same size, stratified to match the
validation split's distribution over dataset source, genre, and number
of active stems. The two external catalogs are used in full.

\paragraph{Context, targets, and prompts.}
Within a section, the stems are divided at random into context and
targets. Each section is first assigned a context size, no context,
one or two stems, or three or more, with probabilities of 0.15, 0.45,
and 0.40, and the context stems are
then drawn at random, and the remaining stems are the targets.
A section with too few stems falls back to a smaller context size,
so the realized contextless share exceeds 0.15
(Table~\ref{tab:composition}). The prompt for each
target is a 1--2\,s excerpt of that stem's own recording in the same
section.

\paragraph{Beat and key.}
For in-house tracks, beat positions and key come from the metadata. For the
external tracks, beats and downbeats are measured by All-In-One and no
key is given.

\paragraph{Generation.}
Each instance is generated once with a fixed seed. A generated stem is
kept when at least 15\% of its 100\,ms frames exceed $-60$\,dBFS, and
dropped from every comparison otherwise, for both systems. ACE
generates the target stems one at a time, rhythm stems first, then
bass, then the rest, mixing each generated stem into the context before
requesting the next; we found this order to give the best results.

\paragraph{Candidate baselines.}
Table~\ref{tab:baselines} lists the systems we considered as
baselines. A baseline must take a per-stem audio prompt and a context,
condition on the prompt audio itself, generate individual stems, and be
runnable, which only ACE-Step v1.5 in Lego mode does.

\begin{table}[H]
\centering\footnotesize  %
\caption{\textbf{Candidate baselines.} Research systems with
released weights (top) and commercial services (bottom). What each
system takes as the prompt for the generated part and
whether it takes existing audio as a context and produces stems (\checkmark{} yes, -- no).}
\label{tab:baselines}
\setlength{\tabcolsep}{2pt}
\begin{tabular}{@{}lllll@{}}
\toprule
System & Prompt & Context & Stems & Why not a baseline \\
\midrule
ACE-Step 1.5, Lego~\citep{acestep2026} & audio, per stem & \checkmark & \checkmark & \emph{baseline} \\
MTMusicLDM~\citep{karchkhadze2024mtmusicldm} & text & \checkmark & \checkmark & no audio prompt \\
MSDM~\citep{mariani2024msdm} & -- & inpainting & 4 fixed & no audio prompt \\
YuE~\citep{yuan2026yue} & audio & -- & 2 fixed & no context \\
MusicGen-Melody~\citep{copet2023musicgen} & chroma & -- & -- & no context \\
Stable Audio Open~\citep{evans2024stableaudioopen} & audio-to-audio & -- & -- & no context \\
\midrule
ACE Studio, Add a Layer & type, text & \checkmark & \checkmark & no audio prompt \\
Google Flow Music & audio$\to$text & song & -- & audio as text \\
Eleven Music & text, lyrics & song & -- & no audio prompt \\
Deezer Remix Lab & presets & song & 3 fixed & genre presets \\
Hook Music & presets & section & -- & preset remix \\
\bottomrule
\end{tabular}

\end{table}

\FloatBarrier

\section{Conditioning Faithfulness Details}
\label{app:faithfulness}

\subsection{Measures}
\label{app:metrics}

This section describes how the values of Tables~\ref{tab:t1}
and~\ref{tab:t3a} are computed.

\paragraph{Table~\ref{tab:t1}: prompt adherence.}
Each value compares one generated stem with one prompt excerpt.
\emph{Matched} uses the stem's own prompt; \emph{Random} uses the prompt
excerpt of a random stem from another track of the same conditioning
source; $\Delta$ is their difference. For style, each clip is resampled
to 24\,kHz and passed through MERT-v1-330M~\citep{li2023mert} in chunks
of at most 5\,s; the hidden states of the last transformer layer are
averaged over time, giving one 1{,}024-dimensional vector per chunk, and
a stem's chunk vectors are averaged into one vector. Raw MERT cosines are
high between any two music clips, so we subtract the mean vector of a
fixed corpus of 1,471 random training segments from every vector before
taking the cosine; unrelated clips then score near zero, and the Random
column reads as a zero point. For timbre, each clip is loudness-normalized
to $-14$\,LUFS and converted to 20 MFCCs per 25\,ms frame, the energy
coefficient dropped, which are averaged over the clip into one vector;
timbre similarity is the cosine between two such vectors, without
centering.

\paragraph{Table~\ref{tab:t3a}: conditioning swaps.}
Timbre
rows use the MFCC similarity of Table~\ref{tab:t1}. Harmony rows use
chroma correlation, the Pearson correlation between 12-bin constant-Q
chromagrams averaged per bar on the beat grid given to the model. For
MFCC, the unrelated pair is a generated stem against another track's
prompt, and the matched reference is a generated
stem against its own prompt. For chroma, the unrelated pair is a
generated stem against another track's context, and the matched
reference is the real stem against its own context.

\subsection{Tempo and Pitch Transformations}
\label{app:transform}

\begin{table}[H]
\centering\small
\caption{\textbf{Tempo and pitch transformations.} One conditioning
input is time-scaled or pitch-shifted by a known amount and the stem
regenerated, while the other conditioning input is held fixed. \emph{Oracle}
is the value perfect following would yield.}
\label{tab:t3b}
\begin{tabular}{llccc}
\toprule
Hypothesis & Metric & Oracle & \Gen & ACE \\
\midrule
\multicolumn{5}{l}{\emph{(a) Prompt time-scaled ($\times$0.9 / 1.1)}}\\
\quad Output tempo unchanged & Tempo accuracy & $100\%$ & $95\%$ & $91\%$\\
\quad Harmony stays with the context & Chroma $\Delta$ & $\approx 0$ & $-0.003$ & $+0.003$\\
\quad Timbre unchanged & MFCC $\Delta$ & $\approx 0$ & $-0.005$ & $-0.005$\\
\midrule\multicolumn{5}{l}{\emph{(b) Prompt pitch-shifted ($\pm 2$\,st)}}\\
\quad Harmony stays with the context & Chroma $\Delta$ & $\approx 0$ & $-0.005$ & $-0.025$\\
\quad Timbre unchanged & MFCC $\Delta$ & $\approx 0$ & $-0.006$ & $-0.006$\\
\midrule\multicolumn{5}{l}{\emph{(c) Context time-scaled ($\times$0.9 / 1.1)}}\\
\quad Output tempo follows & Tempo accuracy & $\approx 90\%$ & $83\%$ & $58\%$\\
\quad Timbre stays with the prompt & MFCC $\Delta$ & $\approx 0$ & $-0.005$ & $-0.008$\\
\midrule\multicolumn{5}{l}{\emph{(d) Context transposed ($\pm 2$\,st)}}\\
\quad Output follows the transposition & Chroma accuracy & $\approx 86\%$ & $76\%$ & $33\%$\\
\quad Timbre stays with the prompt & MFCC $\Delta$ & $\approx 0$ & $-0.001$ & $+0.002$\\
\bottomrule
\end{tabular}

\end{table}

\paragraph{Transformations and hypotheses.}
Table~\ref{tab:t3b} keeps the prompt and context and, instead of
swapping them, applies a simple transformation, a pitch shift or a
time-scale modification, to one of them. As in
Table~\ref{tab:t3a}, only instances that have a context enter this
test.  A moderate transformation of the prompt leaves its
timbre essentially intact, so it should not affect the generation
(a,\,b), whereas a transformation of the context changes its tempo or key, so the
generation must change with it (c,\,d). In (a) the prompt is time-scaled
by $\times$0.9/1.1. The output's tempo should not change, and neither should the chroma correlation between the generation
and the context or the timbre similarity to the prompt. In (b) the prompt
is pitch-shifted by $\pm2$\,st, and the harmony correlation to the context
and the timbre similarity to the prompt should again remain unchanged. In
(c) the context is time-scaled, and the output's tempo should now follow the
applied factor, while
timbre similarity to the prompt must not move. In (d) the context is
transposed by $\pm2$\,st, and the output should follow the transposition,
as measured by chroma accuracy. Perfectly transposed audio scores 86\% on
this measure, its ceiling (the \emph{Oracle} column of the table), and untransposed generations score 14\% (\Gen) and
22\% (ACE), the floor.

\paragraph{MFCC and chroma $\Delta$.}
Rows (a)--(c) compare the regeneration with the base generation of
the same instance. MFCC~$\Delta$ is the timbre similarity of
Table~\ref{tab:t1} between the regenerated stem and the original prompt,
minus the same similarity for the base generation. Chroma~$\Delta$ is the
chroma correlation of Table~\ref{tab:t3a} with the original context,
regeneration minus base, computed on the sum of the generated stems at
the beat level. 

\paragraph{Tempo accuracy.}
Tempo accuracy tests, unit by unit, which tempo the
regeneration took. Both tempi are estimated with librosa's tempo
estimator on the summed generated drum stems, and an estimate that is
off by an octave is folded back. The regeneration's tempo is then compared with two
candidates. The first is the base generation's tempo, which the output
keeps if it ignores the transformation. The second is the tempo the
output would have if it followed the transformation, the base tempo
time-scaled by the same factor. The unit is assigned to the nearer
candidate on a log scale. In (a) the tempo should stay, so tempo
accuracy is the share of units assigned to the first candidate. In (c)
it should follow, so it is the share assigned to the second. The oracle
value for (c) is the base generation itself time-scaled by the same
factor and scored the same way.

\paragraph{Chroma accuracy.}
Row (d) is computed on tonal stems only, drums and percussion excluded, generated once with the original
context and once with the context pitch-shifted by $+2$ and by $-2$
semitones, the key input shifted accordingly. The generation's chroma is correlated as in
Table~\ref{tab:t3a} with the real stem and with the real stem
pitch-shifted by the same amount, and the case counts as followed when
the shifted version correlates more. The ceiling is the base generation
itself pitch-shifted by the same amount and scored the same way, and the
floor is the untransposed base generation.

\paragraph{Results.}
Under the prompt transformations (a,\,b), tempo, harmony, and
timbre stay nearly unchanged for both systems. The output keeps the
base generation's tempo in 95\% of units for \Gen and in 91\% for ACE,
and the chroma and timbre changes stay within 0.03. Under context
time-scaling (c), 
the two part ways. \Gen moves to the new tempo in 83\% of units,
close to the 90\% ceiling, while ACE does so in 58\%, near the 50\% that
a tempo unrelated to the factor would score. The transposition (d)
separates them as well. \Gen follows it with 76\% chroma accuracy,
close to the 86\% ceiling, while ACE reaches 33\%, close to its 22\%
floor. ACE's unchanged harmony correlations in (a,\,b)
may look like stability, but read together with Table~\ref{tab:t3a} and
(d), they reflect the absence of following. ACE's harmony
does not follow the context in the first place, so a prompt
transformation has nothing to disturb.

\section{Memorization Audit}
\label{app:memorization}

\subsection{Audit Protocols in Prior Work}
\label{app:audits}

\noindent\begin{minipage}{\linewidth}
\centering\footnotesize
\vspace{8pt}
\setlength{\tabcolsep}{3pt}
\captionof{table}{\textbf{Memorization audits in prior music generation
papers.} \emph{Search}: whether a generation is compared
against the training set, or only against the source of its own
continuation (\checkmark{} yes, -- no).}
\label{tab:prioraudits}
\begin{tabular}{@{}lcllc@{}}
\toprule
System & Search & Detector & Decision criterion & Listening \\
\midrule
Jukebox \citep{dhariwal2020jukebox} & -- & Listening & Listener judgment & \checkmark \\
MusicLM \citep{agostinelli2023musiclm} & -- & Token match & Calibrated threshold & -- \\
MusicGen \citep{copet2023musicgen} & -- & Token match & Fixed 80\% match & -- \\
Stable Audio 2 \citep{evans2024stableaudio2} & \checkmark & CLAP retrieval & Manual threshold & \checkmark \\
Stable Audio Open \citep{evans2024stableaudioopen} & \checkmark & CLAP retrieval & Manual threshold & \checkmark \\
YuE \citep{yuan2026yue} & \checkmark & ByteCover2 & Distributions only & -- \\
\bottomrule
\end{tabular}
\end{minipage}

Table~\ref{tab:prioraudits} summarizes the six memorization audits
discussed in Section~\ref{sec:related:memaudit}. MusicLM's calibrated
threshold is set on pairs known not to match. Apart from Jukebox,
which relies on listening, and YuE, which uses a cover-song
identification model, their detectors are bound to the sound of the
recording: token sequences encode the rendered audio, so a match requires
near-verbatim reuse, and similarity in general-purpose audio embeddings
such as CLAP is dominated by timbre.

\FloatBarrier
\subsection{Retrieval and Flagging}
\label{app:retrieval}

Each retriever embeds every training stem and every generated stem
in windows and scores a generated stem against each training stem.
\Det and CLEWS~\citep{serra2025clews} use 20\,s windows at a 5\,s
hop, mixed to mono at 16\,kHz, with a stem shorter than 20\,s
repeat-padded to one window; the score is the smallest squared Euclidean
distance between any window of the generated stem and any window of the
training stem, so lower is more similar. CLEWS uses the public
checkpoint trained on SHS100K. CLAP and MERT use the windows of Table~\ref{tab:t1}; the score is the largest centered cosine between any
window of the generated stem and any window of the training stem, so
higher is more similar. A generated stem's score is that of its nearest
training stem.

The search covers every stem of the training split, of every
instrument, not only the melody-carrying stems the queries are drawn
from. For a training query, the exclusion of its own track is applied to
the ranked results by track identifier, so the score comes from the
nearest stem of another track; validation queries need no exclusion,
since their tracks are not in the training set. The thresholds are 0.672
for \Det and 0.812 for CLEWS (distances) and 0.971 for CLAP and
0.974 for MERT (cosines); on the training queries they flag 34, 35, 31,
and 24 stems. The Null band of Figure~\ref{fig:five} is the 5th to 95th
percentile of the same 497 validation scores.

\subsection{Listening Protocol}
\label{app:listening}

\paragraph{Listeners and matches.}
Three listeners judged every flagged stem, following the questions of
Figure~\ref{fig:tree} and the listening procedure of Section~\ref{sec:eval:memorization}.
The three listeners are authors of this paper. The unit of
listening is one match, a flagged stem with one retrieved training stem,
heard together with the full stem the prompt was cut from. A stem that
different retrievers matched to different training stems appears in more
than one match, so the 171 flagged stems give 213 matches. Each match
showed only the audio, the instrument family, and a number. Each listener
heard the matches in a different random order.
\paragraph{Merging.}
Each match's category is the majority vote of the three listeners.
On five of the 213 matches all three differed, and these were settled by
joint listening. The listeners gave the same category on 178 of the
213 matches (84\%, Fleiss' $\kappa=0.75$) and agreed on whether a match
is a memorization on 191 (90\%, $\kappa=0.82$). The one Type~I case was
unanimous.

\paragraph{Borderline cases.}
Three questions of Figure~\ref{fig:tree} raised borderline cases.
For \emph{same composition} (question~1), the listeners set timbre aside
and compared the two stems as note sequences. This was hardest for
synthesizer sounds, where a filter sweep or resonance can read as a
note, and for lines that were nearly identical but differed in a
syncopated note or in the voicing of a chord. For \emph{commonplace}
(question~2), the criterion was whether reusing the shared material in
one's own music would raise a plagiarism concern; since a single stem is
judged rather than a whole piece, this test is stricter than it would be
on a mix. For \emph{duplicate} (question~0), the retrieved stem had to
sound like the same recording as the prompt's full stem, in timbre and
in nearly every note; one or two differing notes were tolerated as a
variant registration of the same stem, but a stem that matched except
for a final chord played as an arpeggio was judged not a duplicate.

\FloatBarrier
\section{Generation Quality}
\label{app:quality}

\begin{table}[H]
\centering\small
\caption{ FAD and listener ratings of
generated stems, and listener ratings of the real stems of the
same cases. FAD: lower is better. Ratings: mean opinion scores from
the listening test (1--5) with 95\% confidence intervals, higher is
better. Quality, Prompt, and Context are sound quality, match to the
prompt audio, and fit with the context.
}
\label{tab:t4}
\begin{tabular}{llc ccc}
\toprule
System & Source & FAD ($\times10^{-3}$)$\downarrow$ & \multicolumn{3}{c}{Human ratings (MOS 1--5)$\uparrow$} \\
\cmidrule(lr){4-6}
 & & & Quality & Prompt & Context \\
\midrule
\multirow{4}{*}{\Gen} & Validation & $0.596$ & $2.81\pm0.29$ & $3.36\pm0.32$ & $3.45\pm0.39$ \\
 & MoisesDB & $0.997$ & $2.77\pm0.33$ & $3.34\pm0.33$ & $3.43\pm0.42$ \\
 & MUSDB18 & $1.010$ & $2.80\pm0.27$ & $3.18\pm0.31$ & $3.20\pm0.30$ \\
\cmidrule(lr){2-6}
 & \emph{All} & $0.868$ & $2.80\pm0.17$ & $3.29\pm0.19$ & $3.36\pm0.22$ \\
\midrule
\multirow{4}{*}{ACE} & Validation & $0.935$ & $2.36\pm0.25$ & $2.73\pm0.34$ & $2.56\pm0.44$ \\
 & MoisesDB & $1.570$ & $2.46\pm0.28$ & $2.62\pm0.33$ & $2.32\pm0.34$ \\
 & MUSDB18 & $1.571$ & $2.02\pm0.26$ & $2.26\pm0.30$ & $2.29\pm0.30$ \\
\cmidrule(lr){2-6}
 & \emph{All} & $1.359$ & $2.28\pm0.16$ & $2.54\pm0.19$ & $2.39\pm0.21$ \\
\midrule
\multirow{4}{*}{Real stems} & Validation & $--$ & $3.65\pm0.25$ & $4.02\pm0.19$ & $3.96\pm0.27$ \\
 & MoisesDB & $--$ & $3.76\pm0.28$ & $4.03\pm0.25$ & $3.81\pm0.23$ \\
 & MUSDB18 & $--$ & $3.65\pm0.20$ & $3.92\pm0.21$ & $3.51\pm0.27$ \\
\cmidrule(lr){2-6}
 & \emph{All} & $--$ & $3.68\pm0.14$ & $3.99\pm0.13$ & $3.76\pm0.15$ \\
\bottomrule
\end{tabular}

\end{table}

\begin{figure}[H]
\centering
\includegraphics[width=0.4\linewidth]{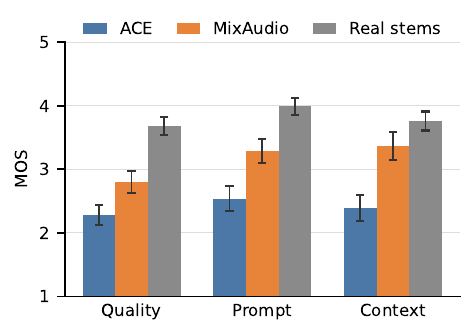}  %
\caption{\textbf{Listener ratings.} Mean opinion scores over
all 48 cases, the \emph{All} rows of Table~\ref{tab:t4}, with their
95\% confidence intervals.}
\label{fig:heval_mos}
\end{figure}

\subsection{Objective Evaluation}
\label{app:quality:fad}
An attribution record matters only if the generated stems are worth
using, so we measure generation quality with Fr\'echet Audio Distance
(FAD)~\citep{kilgour2019frechet} on CLAP-Music
embeddings~\citep{wu2023large}. FAD compares the embedding distribution
of the generated stems with that of real stems held out from the
catalog. For FAD only, each source is split by track: the generated stems of the
evaluation tracks are scored against the real stems of the reference
tracks. The validation split
is divided at random into 100 evaluation and 68 reference tracks (40\%
reference); MoisesDB, which has no official split, into 
70 and 42 (38\%); and MUSDB18 along its official split, the 100 train tracks for
evaluation and the 50 test tracks as reference (33\%). FAD is computed on individual
stems rather than on mixes, and the \emph{All} row averages the three
sources.
Table~\ref{tab:t4} reports both systems on the  three unseen conditioning
sources: \Gen is lower on every source, including the two 
external catalogs.

\subsection{Subjective Evaluation}
\label{app:quality:listening}

\paragraph{Cases.}
The test uses 48 cases, 16 from each of the validation, MoisesDB,
and MUSDB18 sources. A case consists of a prompt excerpt, a context of
one to five stems, the real stem that both systems regenerate,
and, for each system, one generated stem, each heard alone
and mixed with the context. Instances are drawn at random within each
source, and the rated stem is chosen to balance instrument families.
Both systems receive the same conditioning inputs, and playback is
limited to the first 20\,s of the section.

\paragraph{Listeners.}
Seventeen listeners took part, recruited from music programs at
several universities: 11 master's students, 5 undergraduates, and 1
doctoral student, majoring in composition, songwriting, vocal or
instrumental performance, or musicology. Listeners were paid for
their time.

\paragraph{Procedure.}
Each listener rated one practice case and then 16 cases, assigned
so that ratings were spread evenly across cases, each receiving 5.7
ratings on average (4 to 9). A case presents the prompt excerpt, the
context, and three candidates, the real stem and the stem
generated by each system, unlabeled and in random order. Each candidate can be played alone and mixed with the context.
Listeners rate
each candidate from 1 to 5 on sound quality (alone), match to the prompt
in instrument, timbre, and playing style (alone), and fit with the
context in harmony, timing, and groove (mixed). A session took 30
to 40 minutes.

\paragraph{Results.}
Table~\ref{tab:t4} reports the mean opinion score per system and
source, each case first averaged over its listeners, with a 95\%
confidence interval from a bootstrap over cases. \Gen is rated above
ACE on all three axes, and over all cases the intervals do not overlap on any axis.
In Figure~\ref{fig:heval_mos}, the gap between \Gen and the real
stems is smallest on context, and the gap between \Gen and ACE is
largest there. Following the context is thus where \Gen comes closest
to the real stems and gains most over ACE.

\FloatBarrier

\end{document}